\documentclass{webofc}
\usepackage[varg]{txfonts}   
\usepackage{amsmath}
\usepackage{tabu}
\usepackage{nicematrix}
\usepackage{makecell}
\usepackage{multirow}
\usepackage{lineno}

\usepackage{wrapfig}

\makeatletter
\let\LN@equation\equation
\let\LN@endequation\endequation
\renewcommand{\equation}{\linenomath\LN@equation}
\renewcommand{\endequation}{\LN@endequation\endlinenomath}
\let\LN@gather\gather
\let\LN@endgather\endgather
\renewcommand{\gather}{\linenomath\LN@gather}
\renewcommand{\endgather}{\LN@endgather\endlinenomath}
\makeatother

\newcommand{\ns}{{\textsc{ns}}}
\newcommand{\as}{{\textsc{as}}}
\newcommand{\rg}{{\textsc{rg}}}

\newcommand{\pos}{{\textsc{os}}}
\newcommand{\pss}{{\textsc{ss}}}

\newcommand{\nf}{{\text{nf}}}
\begin{document}
%
\title{Estimate of a nonflow baseline for the chiral magnetic effect in isobar collisions at RHIC}
%
%

\author{\firstname{Yicheng} \lastname{Feng}\inst{1}\fnsep\thanks{\email{feng216@purdue.edu}} 
		\firstname{} \lastname{for the STAR Collaboration}
}

\institute{Department of Physics and Astronomy, Purdue University, West Lafayette, IN 47907, USA
          }

\abstract{%
  Recently, STAR reported the isobar (${^{96}_{44}\text{Ru}}+{^{96}_{44}\text{Ru}}$, ${^{96}_{40}\text{Zr}}+{^{96}_{40}\text{Zr}}$) results for the chiral magnetic effect (CME) search.
The Ru+Ru to Zr+Zr ratio of the CME-sensitive observable $\Delta\gamma$, normalized by elliptic anisotropy ($v_{2}$), is observed to be close to the inverse multiplicity ($N$) ratio. In other words, the ratio of the $N\Delta\gamma/v_{2}$ observable is close to the naive background baseline of unity. However, nonflow correlations are expected to cause the baseline to deviate from unity.
To further understand the isobar results,
we decompose the nonflow contributions to $N\Delta\gamma / v_{2}$ (isobar ratio) into three terms
and quantify each term by using the nonflow in $v_{2}$ measurement, published STAR data, and HIJING simulation.
From these studies, we estimate a nonflow baseline of the isobar ratio of $N \Delta\gamma / v_{2}$ for the CME and discuss its implications. 
}
\maketitle
\section{Introduction}
\label{sec:intro}

Quantum chromodynamics (QCD) predicts vacuum fluctuations, rendering nonzero topological charge in a local domain.
As a result, there would be more particles with one certain chirality than the other, 
which is called the chirality anomaly. 
This effect may happen in heavy-ion collisions. 
If there is a strong magnetic field created by the spectator protons in heavy-ion collisions,
the particles would have their spins aligned either parallel or anti-parallel to the magnetic field direction depending on their charges.
Then, with the same chirality preference but opposite spins, the positive and negative charged particles would have opposite momenta.
This charge separation phenomenon is called the chiral magnetic effect (CME).

To search for the CME, STAR conducted the isobar collisions in 2018, 
and recently published their isobar results~\cite{STAR:2021mii}.
The two isobar species, ${^{96}_{44}\text{Ru}}$ and ${^{96}_{40}\text{Zr}}$, were expected to have similar background in CME sensitive observables due to the same nucleon number,
while Ru was expected to have larger CME signal due to more protons and therefore stronger magnetic field.
If so, this CME observable $\Delta\gamma$ over elliptic flow $v_{2}$ would be measured to be larger in Ru+Ru than Zr+Zr.
However, from the recent STAR isobar paper, the Ru+Ru/Zr+Zr ratio of this observable is below unity,
which is on the contrary to the initial expectation.
The main reason is the multiplicity difference between the two isobars.
If we include the multiplicity scaling, namely $N \Delta\gamma /v_{2}$,
then the background baseline would be naively unity. 
The isobar results with the multiplicity scaling are above unity.
In order to conclude whether this is an evidence for CME, one needs to consider other background contributions to $N \Delta\gamma /v_{2}$ from nonflow effects.

\section{$\Delta\gamma$ observable and its nonflow backgrounds}

The CME-sensitive observable 
$\Delta\gamma \equiv C_{3}/v_{2}^{*}$ is defined by
\begin{equation} \label{eq:C3}
\begin{split}
	C_{3,\pos} = \langle \cos(\phi_{\alpha}^{\pm} + \phi_{\beta}^{\mp} - 2\phi_{c}) \rangle, \quad
	C_{3,\pss} = \langle \cos(\phi_{\alpha}^{\pm} + \phi_{\beta}^{\pm} - 2\phi_{c}) \rangle, \quad
	C_{3} = C_{3,\pos} - C_{3,\pss}
	,
\end{split}
\end{equation}
where $\alpha$, $\beta$ indicate particles of interest (POI), 
and $c$ is a reference particle as an estimate of the event plane whose resolution is equal to the elliptic flow $v_{2}^{*}$. 
We use an asterisk ($^{*}$) on $v_{2}$ to indicate that it is the measured $v_{2}$ containing nonflow.
The superscripts $+,-$ indicate the sign of particle charge. 
The subscripts OS, SS stand for opposite-sign and same-sign pairs, respectively. 
Their difference is taken to cancel charge-independent backgrounds.

The $v_{2}^{*}$ measurement contains flow and nonflow, respectively: 
\begin{equation} \label{eq:enf}
	 {v_{2}^{*}}^{2} = v_{2}^{2} + v_{2,\nf}^{2}, \quad \epsilon_{\nf} \equiv v_{2,\nf}^{2} / v_{2}^{2}
	 ,
\end{equation}
where $\epsilon_{\text{nf}}$ denotes their ratio.
$C_{3}$ is composed of a flow-induced background (major), a 
three-particle nonflow correlation (minor), and the possible CME signal~\cite{Feng:2021pgf}. 
Since we mainly focus on the backgrounds, the CME signal term is not written out:
\begin{equation} \label{eq:C3nf}
	C_{3} = \frac{N_{\text{2p}}}{N^{2}} C_{\text{2p}} v_{2,\text{2p}} v_{2} + \frac{N_{\text{3p}}}{2N^{3}} C_{\text{3p}}
	= \frac{v_{2}^{2} \epsilon_{2}}{N} + \frac{\epsilon_{3}}{N^{2}}
	.
\end{equation} 
$C_{\text{2p}} \equiv \langle\cos(\phi_{\alpha}+\phi_{\beta}-2\phi_{\text{2p}}) \rangle$ is the two-particle (2p) nonflow correlations, 
such as  resonance decay daughters  with respect to their  parent azimuthal angle ($\phi_{\text{2p}}$). 
 $ C_{\text{3p}} \equiv \langle\cos(\phi_{\alpha}+\phi_{\beta}-2\phi_{c}) \rangle_{\text{3p}}$ is the three-particle (3p) nonflow correlations,
such as jets where all the three particles are correlated.
Thus, 
\begin{equation} \label{eq:ndgv2nf}
	\frac{N\Delta\gamma}{v_{2}^{*}} = \frac{N C_{3}}{{v_{2}^{*}}^{2}} = \frac{\epsilon_{2}}{1+\epsilon_{\nf}} + \frac{\epsilon_{3}}{N v_{2}^{2} (1 + \epsilon_{\nf})}
	= \frac{\epsilon_{2}}{1+\epsilon_{\nf}} \left( 1 + \frac{\epsilon_{3}/\epsilon_{2}}{N v_{2}^{2}} \right)
	,
\end{equation}
where $\epsilon_{2} \equiv \frac{N_{\text{2p}} v_{2,\text{2p}} }{N v_{2}} C_{\text{2p}}$ and $\epsilon_{3}\equiv \frac{N_{\text{3p}} }{2N} C_{\text{3p}}$ are short-hand notations.
With  approximations to  leading order, we  get the nonflow contributions to the isobar ratio:
\begin{equation} \label{eq:isobarratio}
\begin{split}
	& \frac{(N\Delta\gamma/v_{2}^{*})^{\text{Ru}}}{(N\Delta\gamma/v_{2}^{*})^{\text{Zr}}}
	\equiv \frac{(N C_{3}/{v_{2}^{*}}^{2})^{\text{Ru}}}{(N C_{3}/{v_{2}^{*}}^{2})^{\text{Zr}}} 
	= \frac{\epsilon_{2}^{\text{Ru}}}{\epsilon_{2}^{\text{Zr}}} 
	\cdot \frac{(1+\epsilon_{\nf})^{\text{Zr}}}{(1+\epsilon_{\nf})^{\text{Ru}}}
	\cdot \frac{\left[1 + \epsilon_{3}/\epsilon_{2}/(N v_{2}^{2}) \right]^{\text{Ru}}}{\left[1 + \epsilon_{3}/\epsilon_{2}/(N v_{2}^{2}) \right]^{\text{Zr}}} \\
	\approx & 1 {+ \frac{\Delta \epsilon_{2}}{\epsilon_{2}}}
	{ - \frac{\Delta \epsilon_{\nf}}{1+\epsilon_{\nf}}} 
	{ + \frac{\epsilon_{3}/\epsilon_{2}/(Nv_{2}^{2})}{1 + \epsilon_{3}/\epsilon_{2}/(Nv_{2}^{2})} \left( \frac{\Delta\epsilon_{3}}{\epsilon_{3}} - \frac{\Delta\epsilon_{2}}{\epsilon_{2}} - \frac{\Delta N}{N} - \frac{\Delta v_{2}^{2}}{v_{2}^{2}} \right) }
	.
\end{split}
\end{equation}
In the above expression the quantities with $\Delta$ are the differences between Ru+Ru and Zr+Zr, 
while the others refer to those from Zr+Zr (or similarly Ru+Ru).
We need those $\epsilon$-terms for an improved background estimate.

\section{Nonflow estimates}

\begin{figure}[]
	\includegraphics[width=0.49\linewidth]{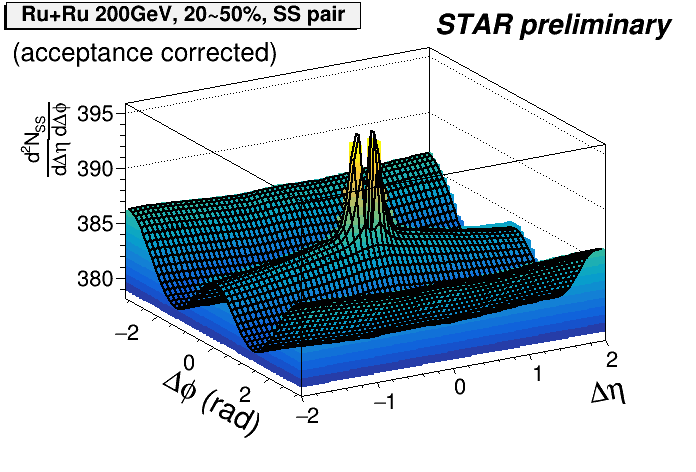}
	\includegraphics[width=0.49\linewidth]{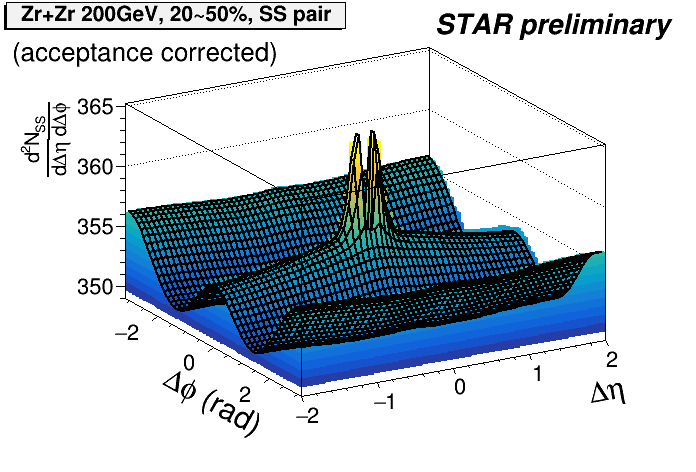}
	\vspace{-0.5cm}
	\caption{The two-particle $(\Delta\eta, \Delta\phi)$ distributions of SS pairs (left: Ru+Ru; right: Zr+Zr). 
	The POI are from $0.2 < p_{T} < 2.0 \text{ GeV}/c$ and $|\eta|<1$. 
	The centrality range is $20$--$50\%$, which is defined by cutting on the POI multiplicity. 
	The acceptance effect is corrected by mixed-event technique.}
	\label{fig:DPhiDEta}
\end{figure}

To get the nonflow in $v_{2}^{*}$ measurement, $\epsilon_{\text{nf}}$, 
we perform fit on the two-particle ($\Delta\eta$, $\Delta\phi$) 2D distribution for mid-central collisions (Fig.~\ref{fig:DPhiDEta}). 
The fit function is given by:
\begin{equation} \label{eq:fitfunc}
\begin{split}
	f(\Delta\eta, \Delta\phi) = &
	 A_{1} G_{\ns,W}(\Delta\eta) G_{\ns,W}(\Delta\phi) 
	+ A_{2} G_{\ns,N}(\Delta\eta) G_{\ns,N}(\Delta\phi) 
	+ A_{3} G_{\ns,D}(\Delta\eta) G_{\ns,D}(\Delta\phi) \\
	&+  \frac{B}{2-|\Delta\eta|} \text{erf}\left( \frac{2-|\Delta\eta|}{\sqrt{2} \sigma_{\Delta\eta,\as}} \right) G_{\as}(\Delta\phi \pm \pi) 
	+  D G_{\rg}(\Delta\eta) \\
	&+  C\big[ 1 + 2 V_{1} \cos(\Delta\phi) + 2 V_{2} \cos(2\Delta\phi) + 2 V_{3} \cos(3\Delta\phi) \big] 
	,
\end{split}
\end{equation}
where $G(x)$ is Gaussian functions. 
The first three terms in Eq.~(\ref{eq:fitfunc}) are 2D Gaussian functions ($A$-terms) for the nearside (NS) correlations. 
The fourth term is for the awayside (AS) correlations ($B$-term), 
whose $\Delta\eta$ projection is an error function divided by ($2-|\Delta\eta|$), and whose $\Delta\phi$ projection is a Gaussian with mean value fixed to $\pm \pi$.
The fifth term ($D$-term)  is a Gaussian centering at $\Delta\eta=0$ and independent of $\Delta\phi$ (referred to as the ridge, RG).
The sixth term is the flow pedestal ($C$-term), where $V_{n} = v_{n}^{2}$ is the squared ``true flows'' assumed to be $\eta$-independent.
Some of the fit results are listed in Table~\ref{tab:FitNonflow}. 
By comparing the ``true flow'' from fit ($V_{2}$) with the inclusive measurement ($\langle \cos(2\Delta\phi) \rangle$ with a $\eta$-gap), we estimate  $\epsilon_{\text{nf}}$ to be approximately  $25\%$. 
The dominant contribution is from the $A_{1}$-term; 
we take half of this term as the systematic uncertainty for $\epsilon_{\nf}$, and obtain 
{$\Delta\epsilon_{\nf}=(-0.82 \pm 0.13 \mp 0.30)\%$}, 
{$-\Delta\epsilon_{\nf}/(1+\epsilon_{\nf}) = (0.65 \pm 0.11 \pm 0.22) \%$}, 
$\Delta v_{2}^{2} / v_{2}^{2} = \Delta V_{2} / V_{2} = (3.7 \pm 0.1 \mp 0.3)\%$.

\begin{table}
\caption{Fit parameters and nonflow calculations.}
\vspace{-0.1cm}
\small
\centering
\begin{tabular}{|c|c|c|c|}
\hline
\multicolumn{2}{|c|}{\bf \it STAR preliminary}  & Ru+Ru & Zr+Zr \\ \hline
\multirow{3}{*}{SS} & fit parameter $C$ & $381.651 \pm 0.011$ & $351.988 \pm 0.009$ \\
  & fit parameter $V_{2} = v_{2}^{2}$ & $0.0029716 \pm 0.0000029$ &  $0.0028668 \pm 0.0000025$ \\
  & $\langle\cos(2\Delta\phi)\rangle_{\pss}$ {\scriptsize $(|\Delta\eta|>0.05)$} & $0.0035968\pm0.0000010$ & $0.0034930\pm0.0000010$ \\
\hline
\multirow{3}{*}{\rotatebox{90}{inclusive}} & $\langle\cos(2\Delta\phi)\rangle = {v_{2}^{*}}^{2}$ {\scriptsize $(|\Delta\eta|>0.05)$} & $0.0037161\pm0.0000007$ & $0.0036088\pm0.0000007$ \\
  & nonflow $U=\langle\cos(2\Delta\phi)\rangle-V_{2}$ & $0.0007446\pm0.0000030$ & $0.0007420\pm0.0000026$ \\
  & $\epsilon_{\nf}=U/V_{2}$ & $(25.06\pm0.10)\%$ & $(25.88\pm0.09)\%$ \\
\hline
\end{tabular}
\label{tab:FitNonflow}
\end{table}


\begin{figure}[b]
	\centering
	\sidecaption
	\includegraphics[width=0.45\linewidth,clip]{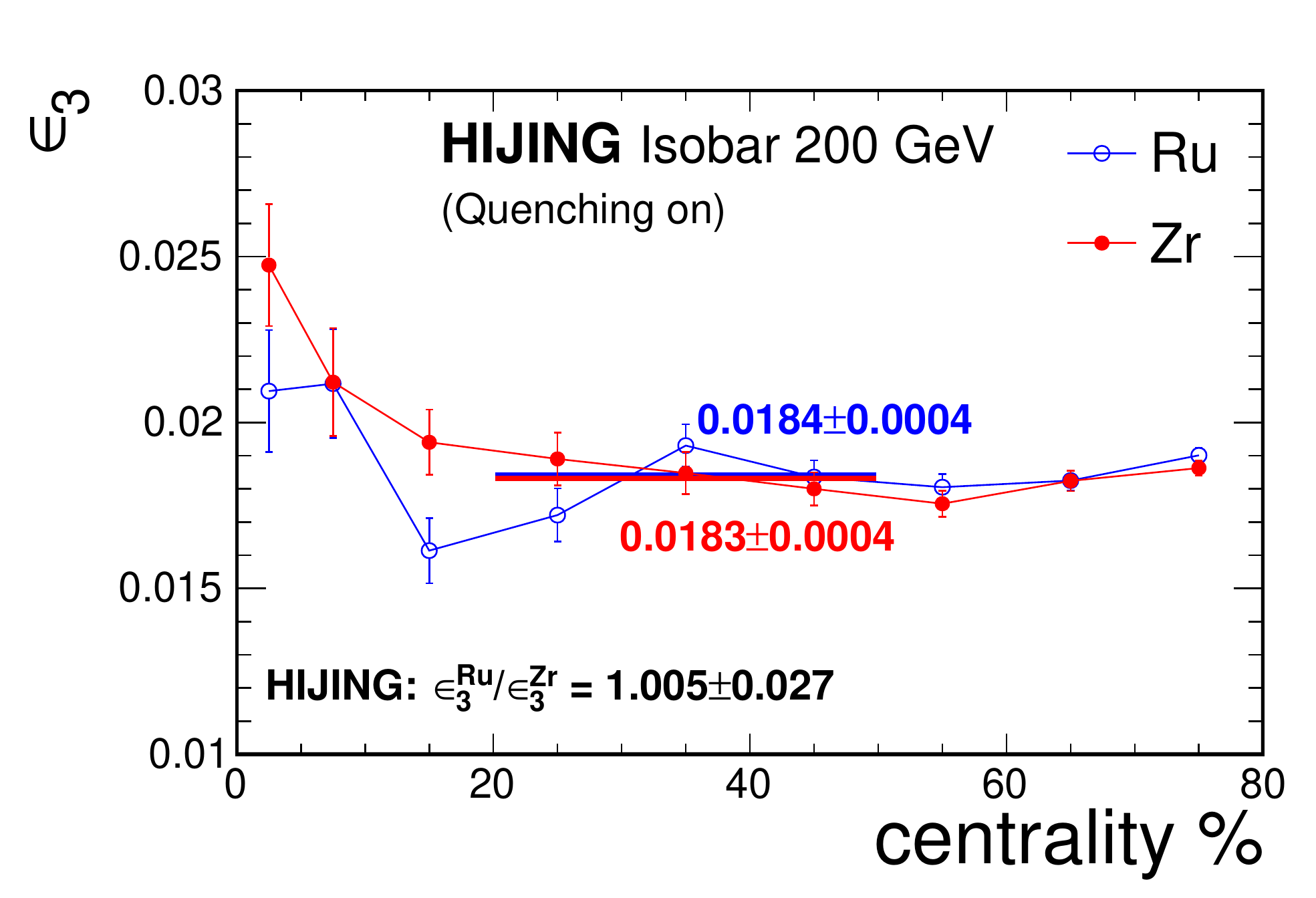}
	\caption{HIJING simulation estimates of $\epsilon_{3}$.}
	\label{fig:Model}
\end{figure}

Due to the large $\eta$ gap between TPC and ZDC, 
there is no 3p nonflow correlation between POI and ZDC event plane, 
so  the $\epsilon_{2}$ can be obtained from ZDC measurements~\cite{STAR:2021mii}: 
$\epsilon_{2} = \frac{N\Delta\gamma\{\text{ZDC}\}}{v_{2}\{\text{ZDC}\}} \approx 0.57 \pm 0.04 \pm 0.02$ (where the tracking efficiency is assumed to be $\sim$80\%).
For the isobar difference, however, the precision of measurements with the ZDC is too poor: $\Delta\epsilon_{2} / \epsilon_{2} \approx (2.3 \pm 9.2)\%$. 
If we assume two isobars have similar $C_{\text{2p}}$, then $\epsilon_{2} \propto N r$,
where the pair multiplicity difference $r \equiv \frac{N_{\pos}-N_{\pss}}{N_{\pos}}$ is precisely measured~\cite{STAR:2021mii}. 
Thus, the isobar difference can be estimated as 
${\Delta\epsilon_{2}/\epsilon_{2} = \Delta r/r + \Delta N/N = (-2.95\pm0.08)\% + 4.4\% = (1.45 \pm 0.08)\%}$, which has better precision.


The 3p nonflow $\epsilon_{3}$ is hard to  estimate from data, 
so we use HIJING model. 
HIJING does not have flow, 
so the only term left should be just the 3p nonflow. 
From this simulation, we get $\epsilon_{3} \approx (1.84 \pm 0.04 \pm 0.92)\%$, 
while the isobar difference is small (consistent with zero) 
$\Delta\epsilon_{3}/\epsilon_{3} = (0.5 \pm 2.7)\%$ (Fig.~\ref{fig:Model}). 
We assumed $50\%$ systematic uncertainty for $\epsilon_{3}$.

\begin{figure}
	\centering
	\includegraphics[width=0.9\linewidth]{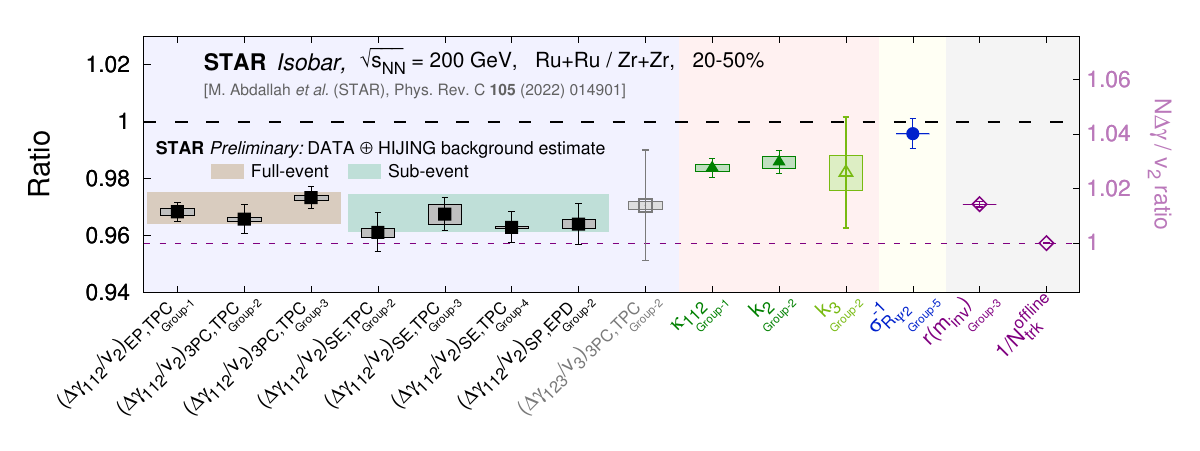}
	\vspace{-0.5cm}
	\caption{
        Background estimates with total uncertainties (bands) on the isobar ratio of $\Delta\gamma/v_{2}$~\cite{STAR:2021mii} 
	.}
	\label{fig:isobar}
\end{figure}

With all the estimates above, we finally get by using Eq.~(\ref{eq:isobarratio}):
\begin{equation}
\begin{split}
	& \frac{(N\Delta\gamma/v_{2}^{*})^{\text{Ru}}}{(N\Delta\gamma/v_{2}^{*})^{\text{Zr}}} 
	\approx  1 {+ \frac{\Delta \epsilon_{2}}{\epsilon_{2}}}
	{ - \frac{\Delta \epsilon_{\nf}}{1+\epsilon_{\nf}}} 
	{ + \frac{\epsilon_{3}/\epsilon_{2}/(Nv_{2}^{2})}{1 + \epsilon_{3}/\epsilon_{2}/(Nv_{2}^{2})} \left( \frac{\Delta\epsilon_{3}}{\epsilon_{3}} - \frac{\Delta\epsilon_{2}}{\epsilon_{2}} - \frac{\Delta N}{N} - \frac{\Delta v_{2}^{2}}{v_{2}^{2}} \right) } \\
	= &
	1 + (1.45\pm0.08)\% + (0.65\pm0.11\pm0.22)\% \\
	&+  (0.094 \pm0.007 \pm0.048) \big[(0.5\pm2.7) - (1.45\pm0.08) - 4.4 - (3.7\pm0.1\pm0.3) \big]\% \\
	= & 1 {+ (1.45\pm0.08)\%} {+ (0.65\pm0.11\pm0.22)\%} { - (0.85 \pm 0.26 \pm 0.44)\%} \\
	=& {1.013 \pm 0.003 \pm 0.005} 
	.
	\label{eq:ratio}
\end{split}
\end{equation}
All the above estimates are for the full-event method~\cite{STAR:2021mii}.
Following the above procedure for sub-events,
we estimate the equivalent of Eq.~(\ref{eq:ratio}) as $(1.011 \pm 0.005 \pm 0.005)$.
We plot those nonflow background estimates together with the STAR isobar data in Fig.~\ref{fig:isobar}.

\section{Summary}

This study estimates nonflow contribution in $v_{2}$ by fitting two-particle $(\Delta\eta,\Delta\phi)$ distribution, and 2p nonflow in $C_{3}$ using STAR isobar data~\cite{STAR:2021mii}. 
The 3p nonflow in $C_{3}$ is evaluated by HIJING simulation. Using the above information, we obtain an improved background estimate of isobar ratio
$\frac{(N\Delta\gamma/v_{2}^{*})^{\text{Ru}}}{(N\Delta\gamma/v_{2}^{*})^{\text{Zr}}} \approx (1.013 \pm 0.003 \pm 0.005)$ for full-event, and $(1.011 \pm 0.005 \pm 0.005)$ for sub-event.


%
%
%

\end{document}